\documentclass[preprint,pre,twocolumn,tightenlines,nobibnotes,10pt]{revtex4}%
\usepackage{amsfonts}
\usepackage{amsmath}
\usepackage{amssymb}
\usepackage{graphicx}%
\newtheorem{theorem}{Theorem}

\newtheorem{remark}[theorem]{Remark}

\begin{document}
\preprint{UATP/04-04}
\title{Complexity Thermodynamics, Equiprobability Principle, Percolation, and
Goldstein's Conjectures}
\author{P. D. Gujrati}
\email{pdg@arjun.physics.uakron.edu}
\affiliation{The Department of Physics, The Department of Polymer Science, The University
of Akron, Akron, OH 44325, USA.}
\date{October 12, 2004}

\pacs{PACS number}

\begin{abstract}
The configurational states as introduced by Goldstein represent the system's
basins and are characterized by their free energies $\varphi(T,V)$ at a given
temperature $T$ and volume $V,$ as we show here. We find that the energies of
some of the special points (termed basin identifiers here) like the basin
minima, maxima, lowest energy barriers, etc., which do not appear in the
partition function so that the latter is independent of them, cannot be used
to characterize the configurational states of the system in all cases due to
their possible non-monotonic behavior as we explain. The complexity
$\mathcal{S(}\varphi,T,V\mathcal{)},$ the natural log of the number of basins
having the same free energy $\varphi(T,V),$ represents the configurational
state entropy. We prove that the equilibrium entropy $S(T,V)$ $\equiv
$\ $\mathcal{S(}\varphi_{\text{b}},T,V\mathcal{)+}S_{\text{b}}(T,V),$\ where
$S_{\text{b}},$ and $\varphi_{\text{b}}$ are the equilibrium basin entropy and
free energy, respectively. We further \ prove that all basins at equilibrium
have the same equilibrium basin energy $E(T,V)$ and entropy $S_{\text{b}%
}(T,V).$ Here, $\varphi$\ and $E$ are measured with respect to the zero of the
potential energy. The equilibrium complexity is shown to be purely an
\ entropic quantity, and not a free energy. The Boltzmann equiprobability
principle is shown to apply to the basins in that each equilibrium basin has
an equal probability $\mathcal{P=}\exp(-\mathcal{S})$ to be explored. This
principle allow us to interpret the basins and their free energy $\varphi$ as
the analogs of the microstates and their energy in the microcanonical
ensemble. In addition, it allows us to draw some useful conclusions about the
time-dependence in the system. We discuss the percolation due to basin
connectivity and its relevance for the dynamic transition. Our analysis
validates modified Goldstein's conjectures that (i) the basin free
energy\ measured with respect to the potential energy of the basin minimum is
insensitive to the basin minimum energy, and (ii) the equilibrium basin free
energy $\varphi_{\text{b}}$\ is insensitive to basins being explored. Thus,
our approach demonstarates that the configurational state description is not
only formally exact, but also analogous to the the conventional approach using
microstates. All the above results are shown to be valid at all temperatures,
and not just low temperatures as originally propsed by Goldstein.

\end{abstract}
\maketitle

\section{Introduction}

In classical statistical mechanics, the Hamiltonian $\mathcal{H}%
(\mathbf{p},\mathbf{q})\equiv K(\mathbf{p})+E(\mathbf{q})$ of a system of $N$
particles in a fixed volume $V$ is a sum of the kinetic energy $K\mathcal{(}%
\mathbf{p)}$ and the potential energy $E(\mathbf{q})$ of the particles; here
$\mathbf{p}$\textbf{\ }and $\mathbf{q}$ represent the collective momenta and
positions of the particles, respectively.\textbf{\ }The dimensionless total
canonical partition function (PF) $\mathcal{Z}_{N}(T,V)$ of the system can be
written as a product of two \emph{independent} integrals%
\begin{equation}
\mathcal{Z}_{N}(T,V)\equiv\left[  \frac{1}{(2\pi\hslash)^{3N}}\int e^{-\beta
K(\mathbf{p})}d\mathbf{p}\right]  \int^{\prime}e^{-\beta E(\mathbf{q}%
)}d\mathbf{q.} \label{TotalPF}%
\end{equation}
Here $d\mathbf{p}$,$d\mathbf{q}$ represent integrations with respect to the
collective momenta and positions of the particles, and $\beta$ the inverse
temperature $1/T$ in the units of the Boltzmann constant $k_{\text{B}}$.

\subsection{Kinetic PF}

The first integral $Z_{N,\text{KE}}(T)$ (including the prefactor) in the
square brackets in (\ref{TotalPF}) contains no interaction energy, and is
trivial to evaluate because of the quadratic nature of $K(\mathbf{p}).$ It is
clearly independent of the volume $V.$ The corresponding free energy
$F_{N,\text{KE}}(T)\equiv-T\ln Z_{N,\text{KE}}(T)$ is the free energy due to
the kinetic energy and is the \emph{same} for \emph{all} systems, regardless
of their potential energy of interaction, and volume. The corresponding
entropy $S_{N,\text{KE}}(T)\equiv-(\partial F_{N,\text{KE}}(T)/\partial
T)$\ and energy $E_{N,\text{KE}}(T)\equiv F_{N,\text{KE}}(T)+TS_{N,\text{KE}%
}(T)$\ are independent of $V,$ and have the same values for all systems$.$%
\ Therefore, we do not have to specifically take into account the contribution
from the kinetic energy part. All we need to do in this case is to add the
contribution $F_{N,\text{KE}}(T)$ to the free energy from the second integral
in (\ref{TotalPF}) at the end of the calculation. This is precisely what we do
here and mostly consider the configurational PF discussed below.

\subsection{Configurational PF}

The dependence on the potential energy (PE) in the PF $\mathcal{Z}_{N}(T,V)$
is through the second integral $Z_{N}(T,V)$ \textbf{(}the prime indicating
integration over \emph{distinct} configurations of the particles),
conventionally called the configurational PF:%
\begin{equation}
Z_{N}(T,V)\equiv\int^{\prime}d\mathbf{q}e^{-\beta E(\mathbf{q})}.
\label{ConfPF}%
\end{equation}
The configurations of the system are represented by phase points in the phase
space $\Gamma_{N}(V)$ defined by the position coordinates $\mathbf{q.}$ Each
phase point represents a microstate of the system. By appending another axis
to $\Gamma_{N}(V)$ to represent the potential energy $E(\mathbf{q})$ of each
point $\mathbf{q},$ we introduce a $3N$-dimensional energy surface $\Sigma
_{N}(V)$ in a ($3N+1)$-dimensional landscape space $\Lambda_{N}(V)$. The
height of the PE surface represents the potential energy $E$ of each phase
point in the space $\Gamma_{N}(V)$. Thus, each configuration or microstate is
characterized by its potential energy $E,$ along with $V$, and $N;$ the
particle number $N$ is kept fixed in this work$.$ For the moment, let us keep
$V$ fixed and consider the slice $d\Sigma_{N}(E,V)$ of the surface $\Sigma
_{N}(V),$ with energies between $E$ and $E+dE$. The slice is given by the
intersection of $\Sigma_{N}(V)$ with the two parallel constant energy surfaces
$\widehat{\Sigma}_{N}(E)$ and $\widehat{\Sigma}_{N}(E+dE)$\ of energy $E\ $and
$E+dE,$ respectively, in $\Lambda_{N}(V).$ This slice defines the
microcanonical ensemble corresponding to energy $E$ and, according to
Boltzmann \cite{Landau}, all configurations or microstate are
\emph{equiprobable} to be explored by the system. In the canonical, i.e., the
$NTV$-ensemble, the entire surface $\Sigma_{N}(V)$ uniquely determines the
thermodynamics of the system, and is therefore of central importance. This
point has been made forcefully in the context of glasses in a recent work
\cite{Ngai} to which we direct the reader.

In equilibrium, only configurations that have their energies $E$ equal to the
equilibrium energy $E(T,V)$ determine the thermodynamics. These are what we
will call the equilibrium configurations at $T$. The equilibration process,
therefore, is highly \emph{selective} and picks out only the equilibrium
configurations that change from one temperature to another; we are of course
neglecting the thermal fluctuations. It is well known \cite{Landau} that the
energy fluctuations $\overline{\Delta E}$ at constant $V,N$\ is given by
$T\surd C_{V,N},$ where $C_{V,N}$ is heat capacity at constant $V,N.$ Thus,
the relative fluctuation $\overline{\Delta E}/E(T,V)$ behaves as 1/$\surd N$
for any finite $T.$ This justifies our neglecting the fluctuations for a
macroscopic system.

In the following, we will only consider a macroscopic system for which $N$,
and $V$ are very large. The thermodynamic limit, which will be implicit in the
following, requires considering $N\rightarrow\infty,V\rightarrow\infty$ such
that $n\equiv N/V$ is kept fixed in the canonical, i.e., the $NTV$-ensemble.
In the related $NTP$-ensemble, we only require $N\rightarrow\infty.$ We will
usually suppress the index $N$ (but continue to show $V)$ in the following for simplicity.

\subsection{Goldstein's Ideas and Basin Identifiers}

We follow Goldstein and consider the $NTV$-ensemble$.$ The PE surface
$\Sigma(V)$ contains many local minima whose importance has been argued by
Goldstein \cite{Goldstein,Goldstein1} for supercooled liquids. He also
provided a qualitative description of the nature of the resulting landscape
created by the distribution of minima and drew attention to the idea of
\emph{configurational states}, which are associated with PE \emph{minima} in
$\Sigma(V).$ The potential wells containing the minima are known as basins or
valleys. The configurational states are collections of allowed configurations
within a basin. Therefore, the equilibrium configurational states must contain
only equilibrium configurations of energy $E=E(T,V).$ However, because of the
importance of the basin minima (BM) at low temperatures, Goldstein has
postulated that the configurational states, instead of being associated with
the entire basins, are characterized by the basin minima (BM), and in
particular by their energy $E_{\text{L}}.$ The BM are not the only point of
interest in basins; the latter have other points of interest\ such as the
lowest energy barrier (LB) or the most probable energy barriers (MPB), i.e.
the barriers with the highest population, which are bound to play an important
role in crossover dynamics or diffusion, as the temperature is raised; this
point is discussed further in the last section. Thus, one can also take the
energy $E^{\text{LB}}$ corresponding LB or $E^{\text{MPB}}$ corresponding to
MPB to characterize the basin depending on the temperature range. One can also
take the highest energy $E^{\text{H}}$ of the basin, the highest barrier (HB)
energy $E^{\text{HB}},$ or the energy of some other special basin point to
characterize the basin. In the following, we will use the term \emph{basin
identifier} to refer to any of the special points of the basins and denote the
corresponding energy by $\mathcal{E}.$

It is obvious that each basin identifier has its usefulness limited to a
certain temperature range. Thus, as the temperature is raised, different
identifiers and their energies become relevant and to focus on only just one
of the identifiers like BM will have a limited applicability. To provide a
description of the system that is useful at all temperatures, we need to focus
on the basin free energy which is determined by the shape and topology of the
basins. This will require a methodology to express the basin free energy in
terms of the equilibrium energies of an identifier in its temperature range.
One of our aims here would be to describe such a methodology.

The shape and topology uniquely determine the free energy of the basins at
\emph{all} temperature, which then uniquely determine the thermodynamics of
the viscous fluid. Because of this, we will describe the configurational
states by specifying the basin free energy so that the characterization
remains valid at all temperatures. Later on, we will see that we can also use
the basin energy to represent the configurational states, because the
equilibrium configurations are identified by the equilibrium energies, as said above.

The main motivation behind the introduction of configurational states appears
to be the ability to express the entropy as a sum of two different
contributions at low temperatures where the system is confined to one of the
basins \cite{Goldstein}. This partitioning requires the two contributions \ to
be independent. Could such a partition and the independence of its two parts
be valid at all temperatures? In his analysis, Goldstein who only considers BM
as the basin identifier to characterize the basin has listed several
conjectures that were common in the field \cite{Goldstein1} at the time and
are expected to be valid at low enough temperatures.

\begin{itemize}
\item[(C1)] \textit{The basin free energy is independent of the possible basin
being explored.}

\item[(C2)] \textit{The basin free energy is particularly independent of the
basin's potential energy minimum }$E_{\text{L}}$\textit{.}

\item[(C3)] \textit{The partition function is a product of the basin and the
BM\ partition functions} $z_{\text{b}}(T,V),$ \textit{and} $Z_{\text{BM}%
}(T,V),$ \textit{respectively}.
\end{itemize}

The BM PF $Z_{\text{BM}}$ \cite{Goldstein1} is defined as%
\begin{equation}
Z_{\text{BM}}(T,V)=\mathbf{\int}N_{\text{BM}}(E_{\text{L}},V)e^{-\beta
E_{\text{L}}}dE_{\text{L}}. \label{GoldsteinConfPF}%
\end{equation}
Here, $N_{\text{BM}}(E_{\text{L}},V)dE_{\text{L}}$ represents the number of
potential energy \emph{minima} in the energy range $E_{\text{L}}$ and
$E_{\text{L}}+dE_{\text{L}}.$ However, the basin PF $z_{\text{b}}$ is left
undefined. From the form of $Z_{\text{BM}}$ in (\ref{GoldsteinConfPF}), it is
most certain that $z_{\text{b}}$ for a given basin is defined by considering
energies measured with respect to the BM $E_{\text{L}}$ of that basin; see
also Schulz \cite{Schulz}. Goldstein emphasizes that basin anharmonicity or
the curvature at its minimum \cite{Angell} may be very important and should be
included in $z_{\text{b}}.$ According to (C1), all equilibrium basins have the
same basin free energy $f_{\text{b}}(T,V)\equiv-T\ln$ $z_{\text{b}}$.
According to (C2), $f_{\text{b}}(T,V)$ is independent of PE minima of the
equilibrium basins. This conjecture merely reflects the fact that
$f_{\text{b}}(T,V)$\ is measured with respect to the basin energy minimum
\cite{Gujrati}. According to (C3), the PF [without $Z_{\text{KE}}(T)$] is
given as a product \cite{Goldstein1,Schulz}\ \ \ \ \ \ \ \ \ \ \ \ \ \ \ \
\begin{equation}
Z(T,V)=z_{\text{b}}(T,V)Z_{\text{BM}}(T,V). \label{GoldsteinPart}%
\end{equation}
The form of $Z_{\text{BM}}$ in (\ref{GoldsteinConfPF}) implies the
\emph{equiprobability hypothesis} that basins having their BM at the same
$E_{\text{L}}$ are equiprobable. Can these conjectures be justified at all temperatures?

Our aim in this work is to provide a unified formalism that remains valid at
any temperature. As we will see, this will require modifying the above
conjectures slightly, which can then be shown to be valid at all temperatures.

\subsubsection{Condition of Equilibrium and T-Dependence}

\begin{remark}
For a macroscopic system in equilibrium, the PF must be \emph{maximized}.
\end{remark}

We apply this condition to $Z_{\text{BM}}(T,V).$ In equilibrium at a given
$T,V$, it should be dominated by basins having their minima at the equilibrium
minima energy $E_{\text{L}}(T,V)$\ at which the integrand in
(\ref{GoldsteinConfPF}) is maximum. In terms of the BM entropy $S_{\text{BM}%
}(E_{\text{L}},V)\equiv\ln N_{\text{BM}}(E_{\text{L}},V),$\ $E_{\text{L}%
}(T,V)$ is determined by
\begin{equation}
\left[  \left(  \partial S_{\text{BM}}(E_{\text{L}},V)/\partial E_{\text{L}%
}\right)  _{V,N}\right]  _{E_{\text{L}}=E_{\text{L}}(T,V)\ }=\beta,
\label{Goldstein1}%
\end{equation}
which is the condition of equilibrium. The equilibrium BM entropy
$S_{\text{BM}}(T,V)$\ in the canonical ensemble is the value of the
microcanonical entropy $S_{\text{BM}}(E_{\text{L}},V)$\ at $E_{\text{L}%
}=E_{\text{L}}(T,V).$ Even though $S_{\text{BM}}(E_{\text{L}},V)$ is
independent of $T,$ the canonical entropy $S_{\text{BM}}(T,V)\equiv
$\ $S_{\text{BM}}[E_{\text{L}}(T,V),V],$ obtained after the integrand has been
maximized to achieve equilibrium and $E_{\text{L}}$ replaced by its
equilibrium value $E_{\text{L}}(T,V),$ is a function of $T$. This is a common
feature in statistical mechanics.

\begin{remark}
A microcanonical quantity is independent of $T$ but becomes $T$-dependent in
the canonical ensemble through its dependence on some equilibrium quantity
such as the $T$-dependence of $S_{\text{BM}}(E_{\text{L}},V)$\ at
$E_{\text{L}}=E_{\text{L}}(T,V)$ above.
\end{remark}

\subsubsection{Basins at Equilibrium}

Provided (C3) is valid, the basins that contribute at equilibrium have the
same equilibrium BM energy $E_{\text{L}}=E_{\text{L}}(T,V)$ and are
\emph{equally probable}. The BM free energy $F_{\text{BM}}(T,V)\equiv-T\ln
Z_{\text{BM}}(T,V)$ is given by
\[
F_{\text{BM}}(T,V)=E_{\text{L}}(T,V)-TS_{\text{BM}}(T,V),
\]
and the configurational free energy $F\mathcal{(}T,V)\equiv-T\ln Z(T,V)$ is
given by the sum%
\begin{align*}
F\mathcal{(}T,V)  &  =f_{\text{b}}(T,V)+F_{\text{BM}}(T,V)=\\
&  [e_{\text{b}}(T,V)+E_{\text{L}}(T,V)]-T[S_{\text{b}}(T,V)+S_{\text{BM}%
}(T,V)],
\end{align*}
where $e_{\text{b}}(T,V)$ and $S_{\text{b}}(T,V)$ are the equilibrium basin
energy (measured with respect to the basin minima) and entropy, respectively.
The first quantity, therefore, represents the equilibrium basin energy
measured with respect to the zero potential. Adding to this $F_{\text{KE}%
}(T)\equiv E_{\text{KE}}(T)-TS_{\text{KE}}(T)$ from the kinetic PF$,$ we
obtain the free energy%
\begin{subequations}
\begin{align*}
\mathcal{F}(T,V)  &  \equiv-T\ln\mathcal{Z}(T,V)\\
&  =E_{\text{vib}}(T,V)-T[S_{\text{vib}}(T,V)+S_{\text{BM}}(T,V)],
\end{align*}
where $E_{\text{vib}}(T,V)=E_{\text{KE}}(T)+e_{\text{b}}(T,V)+E_{\text{L}%
}(T,V),$ and $S_{\text{vib}}(T,V)=S_{\text{KE}}(T)+S_{\text{b}}(T,V)$
represent the equilibrium basin vibrational energy and entropy, respectively.

Within the approximation implied in (C3) which is expected to be valid at low
temperatures, we observe that the equilibrium entropy $S(T)$ in the above free
energy $\mathcal{F}(T,V)$ or $F\mathcal{(}T,V)$ is a sum of two parts, one of
which is the configurational state entropy $S_{\text{BM}}(T,V),$ and the other
one is $S_{\text{vib}}(T,V)$ or $S_{\text{b}}(T,V),$ respectively. Thus, the
approximation in (C3) and the entropy additivity are equivalent in the above approach.

\subsection{A Different Approach \qquad}

According to Goldstein, we can interpret the thermodynamics of the system as
being confined to any of the many basins; this is certainly expected to be
valid at least at low temperatures. His seminal ideas based on the above
conjectures (C1-C3) have been central in understanding glasses or amorphous
materials, as they provides an intuitive rational for the observed relaxation
of two different time-scales \cite{Goldstein,Johari}. Using these ideas,
Goldstein was able to provide an interesting and sufficiently tractable scheme
for calculation discussed above, and made some quantitative predictions
\cite{Goldstein,Goldstein1}. Our interest here is to check the validity of the
above conjectures, to see if they are valid at all temperatures, and to shed
new light on some of the core concepts, like the basin free energy, the nature
and significance of configurational states and their relationship with the
basin energy or free energy, the \textit{\`{a} priory }equiprobability of the
basins in equilibrium and their energies and entropies, etc. We also discuss
the phenomenon of percolation of basin connectivity and its possible
connection with the dynamics in the system.

We carry out an analysis by taking an approach which has appeared in the
literature \cite{Parisi,Parisi1,Coluzzi,Coniglio}, but has not been studied
extensively. In particular, we measure the basin free energy with respect to a
\emph{common reference energy, }the zero of the potential energy\emph{ }(see
below) and denote it by $\varphi_{\text{b}}(T,V)$ $\ $to distinguish it from
the previous free energy $f_{\text{b}}(T,V).$ The approach leads to the
equiprobability theorem in a straightforward manner. This is highly pleasing
as it brings forth the similarity with the conventional equiprobability
hypothesis due to Boltzmann \cite{Landau}. We substantiate the first two
conjectures of Goldstein in slightly different forms; the third conjecture
also needs to be modified. All these are shown to be valid at \emph{all} temperatures.

\subsubsection{Complexity}

As said above, Goldstein's ideas seem to justify a well-accepted belief in the
glass community that the entropy of the supercooled liquid is a sum of two
contributions: (i) the entropy $S_{\text{BM}}(T,V)$ due to different
equilibrium configurational states or basins, and (ii) the entropy
$S_{\text{b}}(T,V)$ associated with an equilibrium basin. (The supercooled
liquid is a metastable liquid, which is described by the landscape in which
some of the basins that are relevant for the crystal phase are deleted from
consideration.) However, whether such a partition actually occurs in viscous
liquids has never been rigorously demonstrated. (As we will see here, the
actual rigorous demonstration is not trivial.)\ Borrowing ideas from the
spin-glass field, M\'{e}zard and Parisi \cite{Parisi} introduced the idea of
\emph{complexity} $\overline{\mathcal{S}}(T,V)$\ as a candidate for the
entropy of equilibrium configurational states, which has been adopted by other
workers in the field; see, for example, \cite{Coluzzi,Coniglio}. The
complexity is determined not by the BM energies $E_{\text{L}}$ (or the
energies of other basin identifiers such as $E^{\text{LB}},E^{\text{MPB}},$
etc.) but by the basin free energies; see below. Coluzzi et al \cite{Parisi1}
gave a plausibility argument that the liquid's equilibrium entropy $S(T,V)$ is
a sum of the complexity $\overline{\mathcal{S}}(T,V)$ and $S_{\text{b}}(T,V):$%
\end{subequations}
\begin{equation}
S(T,V)=\overline{\mathcal{S}}(T,V)+S_{\text{b}}(T,V). \label{EntropyPartition}%
\end{equation}
Coluzzi and coworkers \cite{Coluzzi} have given a numerical scheme to evaluate
complexity directly by making several hypotheses, one of which is the
equiprobability of the basins of the same free energy, and another one is the
above entropy partition (\ref{EntropyPartition}).

\subsubsection{Present Goal}

Since it appears, from all accounts, that the central idea of configurational
states or basins and the complexity are important concepts that play a pivotal
role in the thermodynamics at low temperatures and hopefully help us unravel
the mystery of glassy states, it is also important to demonstrate that the
above partition in (\ref{EntropyPartition}) is valid rigorously and that the
equiprobability hypothesis can be justified on general grounds not only at low
temperatures but at all temperatures. The latter allows us to treat the basin
free energy $\varphi(T,V)$ in the same manner as the energy in the
microcanonical ensemble because of the equiprobability of configurations of
the same energy in that ensemble. The partitioning of $S(T,V)$ replaces the
third conjecture of Goldstein.

We will prove the following statements.

\begin{itemize}
\item[(S1)] \textit{The basin free energy }$f_{j}(T,V)$\textit{\ of the }%
$j$-th\textit{ basin is independent of the energy of a particular basin
identifier, like the BM energy.}

\item[(S2)] \textit{The equilibrium basin free energy }$\varphi_{\text{b}%
}(T,V)$\textit{, energy }$E_{\text{b}}(T,V),$\textit{\ and entropy
}$S_{\text{b}}(T,V)$\textit{\ are the same in all equilibrium basins that are
explored at a given }$T,V$\textit{. }

\item[(S3)] \textit{All equilibrium basins have equal probability of being
sampled.}

\item[(S4)] \textit{The entropy partition }(\textit{\ref{EntropyPartition}%
})\textit{\ is a general result.}
\end{itemize}

We discuss the general topological features of the landscape in the next
section, and write down the fundamental thermodynamic differential relations.
We then compare this analysis with the standard analysis, which is carried out
in Sect. III. The results are discussed in the last section, which also
contains some speculative comments and a short summary of our results.

\section{Goldstein's Landscape}

\subsection{Analysis}

The surface $\Sigma(V)$ can be formally decomposed into nonoverlapping basins
$B_{j}(V).$ There would be an energy barrier between different neighboring
minima. In each basin $B_{j}(V),$ the allowed energy range is $\Delta
_{j}E=E_{j}^{\text{H}}-E_{\text{L}j},$ where $E_{j}^{\text{H}}$ and
$E_{\text{L}j}(V)$ are the highest and lowest allowed energies in $B_{j}(V)$
\cite{Note}$.$ The highest energy also represents the highest barrier energy
in the basin. The exceptions are the basins that contain particles near the
walls of the container (of volume $V$) in which the interactions of the
particles with the walls will make the potential very large ($\rightarrow
\infty$) in order to confine the particles. For such basins, $E_{j}^{\text{H}%
}\rightarrow\infty.$ In this case, the highest energy does not represent the
highest barrier to some other basin. The lowest energy in a basin does not
necessarily have to be point-like in that it is associated with only one phase
space point $\mathbf{q}$. It is possible that there are many contiguous phase
space points, all having the \emph{same} lowest energy, with no energy barrier
between them such as the minima in an inverted Mexican hat. In this case, all
these points would be classified as representing a \emph{single} minimum and
belonging to a single basin. The same situation may also occur for other
points like the points of highest energy, the energy barriers, etc. The number
of basins in $\Gamma(V)$ will be denoted by $N_{\text{B}}(V).$ The canonical
partition function $Z(T,V)$ can be formally expressed in the form of an
identity as a sum over the basins:%
\begin{equation}
Z(T,V)\equiv\sum_{j=1,N_{\text{B}}(V)}Z_{j}(T,V), \label{PF}%
\end{equation}
where
\begin{equation}
Z_{j}(T,V)\equiv%
{\displaystyle\int\limits_{B_{j}(V)}}
d\mathbf{q}e^{-\beta E(\mathbf{q})} \label{BasinPF}%
\end{equation}
is the PF of the $j$-th basin $B_{j}(V)$, and the integration is over the
coordinates confined within the basin. The energy is measured with respect to
a \emph{common} zero of potential for all basins, and not with respect to some
predetermined $\mathcal{E}_{j}(V)$ associated with a basin identifier. We
denote the basin free energy by $\varphi_{j}(T,V)\equiv-T\ln Z_{j}(T,V)$ and
introduce $f_{j}(T,V)\equiv\varphi_{j}(T,V)-\mathcal{E}_{j}(V),$which
represents the basin free energy measured from $\mathcal{E}_{j}(V).$ The
probability $\mathcal{P}_{j}(T,V)$ that the system under equilibrium explores
the basin $B_{j}(V)$ is given by%
\begin{equation}
\mathcal{P}_{j}(T,V)\equiv Z_{j}(T,V)/Z(T,V). \label{Prob}%
\end{equation}

\subsection{\bigskip Fundamental Thermodynamic Relations}

\subsubsection{Basins}

Let us rewrite (\ref{BasinPF}) in a different way as follows. Consider the
surface $\Sigma_{j}(V),$ which is the part of the surface $\Sigma(V)$ that
belongs to the basin $B_{j}(V).$ The surface $\Sigma_{j}(V)$ exists for
energies in the range $\Delta_{j}E$. We project it on the 3$N$-dimensional
phase space $\Gamma(V)$, which is orthogonal to the $E$-axis. The projection
$\Pi_{j}(V)$\ determines the number of configurations (not to be confused with
the configurational states of Goldstein) of the system that belong to the
basin as follows. Consider the slice $d\Sigma_{j}(E,V)$\ of the surface
$\Sigma_{j}(V)$\ between the energies $E$ and $E+dE,$~$E\in\Delta_{j}E.$ The
slice contains all the points in $\Sigma_{j}(V)$ that lie between the
intersections (known as the \emph{turning points} in classical mechanics) of
$\Sigma_{j}(V)$ with the two parallel constant energy surfaces $\widehat
{\Sigma}(E)$ and $\widehat{\Sigma}(E+dE)$ that were introduced above, and will
be \emph{closed} (in the sense that a ring is) if the slice cuts the basin in
all directions; otherwise it is \emph{open}$.$ For energies close to
$E_{\text{L}j}(V),$ the slice is closed, but becomes open as we get to higher
energies and eventually disappears at\ $E_{j}^{\text{H}}(V).$ The projection
$d\Pi_{j}(E,V)$ of this slice on $\Gamma(V)$ is closed (open) if $d\Sigma
_{j}(E,V)$ is closed(open).\ The area $\left\vert d\Pi_{j}(E,V)\right\vert $
of the projection determines the number of configurations $W_{j}(E,V)dE=$
$\left\vert d\Pi_{j}(E,V)\right\vert $ between the energies $E$ and $E+dE.$
Let us introduce the microcanonical \emph{basin} entropy function
$S_{j}(E,V)\equiv\ln W_{j}(E,V),$ which is independent of $T.$ The entropy
$S_{j}(E,V)$ at fixed $V,N$\ satisfies the differential relation%
\begin{equation}
dS_{j}(E,V)=\left(  \partial S_{j}(E,V)/\partial E\right)  _{V,N}dE.
\label{DifferentialS_j}%
\end{equation}
Now, we can rewrite (\ref{BasinPF}) as follows:
\[
Z_{j}(T,V)\equiv%
{\displaystyle\int\limits_{E\in\Delta_{j}E}}
dE\exp[S_{j}(E,V)-\beta E(V)].
\]
For a macroscopic system at fixed $T,$ the integral is dominated by the
maximum integrand at $E=E_{j}(T,V)$ (see Remark 1), given by
\begin{equation}
\left[  \left(  \partial S_{j}(E,V)/\partial E\right)  _{V,N}\right]
_{E=E_{j}(T,V)}=\beta, \label{BasinSTrelation}%
\end{equation}
which is the standard thermodynamic relation between the entropy and
temperature; compare with (\ref{Goldstein1}). The energy\ $E_{j}%
(T,V)$\ represents the \emph{average} basin energy at a given temperature
$T$,$V,$ (and $N).$ (We will reserve the usage of the term "equilibrium" for a
thermodynamic quantity only after the PF $Z(T,V)$ has been evaluated. To make
this distinction, we use the term "average" at this stage.) Let us introduce
the energy difference $\Delta E_{j}=E-\mathcal{E}_{j},$ and $\Delta
E_{j}(T,V)=E_{j}(T,V)-\mathcal{E}_{j}$\ for the $j$-th basin, where
$\mathcal{E}_{j}$ is the energy of some particular basin identifier like BM,
LB, etc. We can rewrite $\left(  \ref{BasinSTrelation}\right)  $ as
\begin{equation}
\left[  \left(  \partial S_{j}(E,V)/\partial\Delta E_{j}\right)
_{V,N}\right]  _{\Delta E_{j}=\Delta E_{j}(T,V)}=\beta.
\label{BasinSTrelation1}%
\end{equation}
The \emph{canonical }basin entropy $S_{j}(T,V)$\ is given by the
microcanonical basin entropy evaluated at the average $E=E_{j}(T,V):$%
\[
S_{j}(T,V)\equiv S_{j}[E_{j}(T,V),V].
\]
Using (\ref{DifferentialS_j}) at $E=E_{j}(T,V),$ and (\ref{BasinSTrelation}),
we find that
\begin{equation}
dE_{j}(T,V)=TdS_{j}(T,V) \label{DifferentialE_j}%
\end{equation}
at fixed $V,N.$ The average basin free energies $\varphi_{j}(T,V)$ and
$f_{j}(T,V)\equiv$ $\varphi_{j}(T,V)-\mathcal{E}_{j}$ are given by%
\begin{subequations}
\begin{align}
\varphi_{j}(T,V)  &  =E_{j}(T,V)-TS_{j}(T,V),\label{IntrabasinFREE}\\
f_{j}(T,V)  &  =\Delta E_{j}(T,V)-TS_{j}(T,V), \label{IntrabasinFREE2}%
\end{align}
A given $T$ corresponds to a particular value of $\Delta E_{j}(T,V)$ [see
$\left(  \ref{BasinSTrelation1}\right)  ]$ measured from $\mathcal{E}_{j}.$
Thus, $\Delta E_{j}(T,V)$ will remain the same even if the value of
$\mathcal{E}_{j}$ is changed by shifting the origin of the energy. In other
words, $\Delta E_{j}(T,V)$ and the average basin free energy $f_{j}(T,V)$ do
not depend on $\mathcal{E}_{j}.$\ This proves our statement (S1).
Consequently, $\varphi_{j}(T,V)\equiv$\ $f_{j}(T,V)$ $+$ $\mathcal{E}_{j}$
depends \emph{linearly} on $\mathcal{E}_{j}.$ Using (\ref{DifferentialS_j}),
(\ref{BasinSTrelation}), and (\ref{DifferentialE_j}), it is easy to obtain the
following fundamental differential relation at constant $V$,$N:$
\end{subequations}
\begin{equation}
d\varphi_{j}(T,V)=df_{j}(T,V)=-S_{j}(T,V)dT. \label{IntrabasinChange}%
\end{equation}
This gives the change in the basin free energy due to the temperature change
within the basin $B_{j}(V).$

\subsubsection{Complexity Analysis}

At a given $T,V$ (and $N$), the basin free energy $\varphi_{j}(T,V)$\ varies
from basin to basin; hence, $\varphi_{j}(T,V)$\ represents a \emph{family} of
functions, one for each $j$. Let $\varphi\equiv\varphi(T,V)$ represent one of
the various functions $\varphi_{j}\prime s$ in this family$,$ and
$\mathcal{N}(\varphi,T,V)$ the number of basins having the same free energy
$\varphi$ for a given $T,V,$ regardless of whether $\mathcal{E}_{j}$ is the
same or not for these basins. In other words, these basins need not all have
the same basin identifier energy $\mathcal{E}_{j}=\mathcal{E}.$ We now rewrite
(\ref{PF}) as%
\begin{equation}
Z(T,V)\equiv\sum_{\varphi}\mathcal{N}(\varphi,T,V)e^{-\beta\varphi}.
\label{PF1}%
\end{equation}
For a macroscopic system, we expect the family of functions represented by
$\varphi(T,V)$ to be almost \emph{continuous} "over" the basin index $j.$ This
is an assumption that is expected to remain valid even at low enough
temperatures. With this assumption, the sum in (\ref{PF1}) can be replaced by
an integration over this family.

Let us introduce the \emph{complexity} $\mathcal{S}(\varphi,T,V)\equiv
\ln\mathcal{N}(\varphi,T,V),$ which satisfies the following differential
relation at constant $V$ (and $N):$%
\begin{equation}
d\mathcal{S}(\varphi,T,V)=\left(  \partial\mathcal{S}/\partial\varphi\right)
_{T,V}d\varphi+\left(  \partial\mathcal{S}/\partial T\right)  _{\varphi,V}dT.
\label{DifferentialS}%
\end{equation}
For a macroscopic system at fixed $T,V,$ see Remark 1, the PF in (\ref{PF1})
is dominated by that particular value of the basin free energy, which we
denote by $\varphi_{\text{b}}\equiv\varphi_{\text{b}}(T,V),$ for which the
summand (or integrand) is maximum, i.e. $\Phi(\varphi,T,V)\equiv
\varphi-T\mathcal{S}$ is minimum. The basins having this free energy
$\varphi_{\text{b}}(T,V)$ represent the \emph{equilibrium basins} determining
the thermodynamics at fixed $T,V.$ As said above, these equilibrium basins are
not required to have the same basin identifier energy $\mathcal{E}%
_{j}=\mathcal{E}.$ The location of the maximum at $\varphi=\varphi_{\text{b}%
},$ for continuous $\varphi,$ is determined by%
\begin{equation}
\left[  \left(  \partial\mathcal{S}(\varphi,T,V)/\partial\varphi\right)
_{T,V}\right]  _{\varphi=\varphi_{\text{b}}(T,V)}=\beta,
\label{Equilibriumphi}%
\end{equation}
where the derivative is obtained by a variation over the family of function
$\varphi$. The temperature $T$ is kept fixed under this variation and is shown
explicitly in (\ref{Equilibriumphi}); this was not necessary in
(\ref{BasinSTrelation}) since $S_{j}(E,V)$ did not have an $T$-dependence,
even though the variation there was also carried out at fixed $T$. The
equilibrium complexity $\overline{\mathcal{S}}(T,V)$ is given by
$\mathcal{S}(\varphi,T,V)$ evaluated at $\varphi=\varphi_{\text{b}}(T,V):$
\begin{equation}
\overline{\mathcal{S}}(T,V)\equiv\mathcal{S}[\varphi_{\text{b}}(T,V),T,V].
\label{InterbasinEntropy}%
\end{equation}
(Here, we have used an overbar to indicate that $\overline{\mathcal{S}}(T,V)$
is a function of two variables, while $\mathcal{S}[\varphi_{\text{b}%
}(T,V),T,V]$ is a function of three variables.) Each equilibrium basin has the
same free energy $\varphi_{\text{b}}.$ This proves a part of our statement (S2).

\subsubsection{Basin Entropy and Complexity\qquad}

We can invert the function $\mathcal{S}(\varphi,T,V)$ over the entire positive
temperature range where it is monotonically increasing with $\varphi,$ see
(\ref{Equilibriumphi}), and express $\varphi$ as a function of $\mathcal{S}%
:\varphi\equiv\varphi(\mathcal{S},T,V)$ over this range$.$\ Accordingly, the
change in $\varphi$ due to changes in the complexity $\mathcal{S}$ and $T$ is
given by
\begin{equation}
d\varphi(\mathcal{S},T,V)=\left(  \partial\varphi/\partial\mathcal{S}\right)
_{T,V}d\mathcal{S+}\left(  \partial\varphi/\partial T\right)  _{\mathcal{S}%
,V}dT. \label{DifferentialPhi}%
\end{equation}
In equilibrium, $\varphi$ and $\mathcal{S}$\ should be replaced by
$\varphi_{\text{b}}$ and $\overline{\mathcal{S}},$ respectively, and the
coefficient of $d\overline{\mathcal{S}},$\ according to (\ref{Equilibriumphi}%
), should be $T$. From (\ref{DifferentialS}), we note that at equilibrium
\[
\left(  \partial\varphi_{\text{b}}(T,V)/\partial T\right)  _{\overline
{\mathcal{S}},V}=-T\left(  \partial\overline{\mathcal{S}}(T,V)/\partial
T\right)  _{\varphi_{\text{b}},V}.
\]
In the identity%
\begin{equation}
\left(  \frac{\partial\varphi_{\text{b}}}{\partial T}\right)  _{V}%
\equiv\left(  \frac{\partial\varphi_{\text{b}}}{\partial T}\right)
_{\overline{\mathcal{S}},V}+\left(  \frac{\partial\varphi_{\text{b}}}%
{\partial\overline{\mathcal{S}}}\right)  _{T,V}\left(  \frac{\partial
\overline{\mathcal{S}}}{\partial T}\right)  _{V},\nonumber
\end{equation}
obtained from (\ref{DifferentialPhi}), the first term on the right represents
the change in the equilibrium basin free energy, see (\ref{IntrabasinChange}),
solely due to the temperature change $dT$ within a basin; the contribution
from the complexity change due to $dT$ is given by the second term. Thus, the
first term is the negative of the basin entropy $S_{\text{b}}(T,V),$ and we
have%
\begin{subequations}
\begin{align}
T\left(  \frac{\partial\overline{\mathcal{S}}}{\partial T}\right)
_{\varphi_{\text{b}},V}  &  =S_{\text{b}}%
(T,V),\label{Basin_Complexity_Relation1}\\
\left(  \frac{\partial\varphi_{\text{b}}}{\partial T}\right)  _{V}  &  \equiv
T\left(  \frac{\partial\overline{\mathcal{S}}}{\partial T}\right)
_{V}-S_{\text{b}}(T,V). \label{Basin_Complexity_Relation2}%
\end{align}

\subsubsection{Free Energy and Entropy\qquad}

The equilibrium free energy $F(T,V)\equiv-T\ln Z(T,V)$\ is given by the value
of $\Phi(\varphi,T,V)\equiv\varphi-T\mathcal{S}$ at equilibrium, i.e.,
\end{subequations}
\begin{equation}
F(T,V)\equiv\Phi(\varphi_{\text{b}},T,V)=\varphi_{\text{b}}(T,V)-T\overline
{\mathcal{S}}(T,V). \label{FreeEnergy0}%
\end{equation}
Taking its differential at constant $V,($and $N)$, we obtain
\begin{align}
dF(T,V)  &  =d\varphi_{\text{b}}(T,V)-Td\overline{\mathcal{S}}(T,V)-\overline
{\mathcal{S}}(T,V)dT\nonumber\\
&  =-[\mathcal{S}(T,V)+S_{\text{b}}(T,V)]dT, \label{DiffF2}%
\end{align}
where we have used (\ref{DifferentialPhi}) applied to\ $\varphi_{\text{b}}$.
The entropy can be calculated directly from $S(T,V)=-(\partial F/\partial
T)_{V}.$ From (\ref{DiffF2}), we have
\[
S(T,V)=\overline{\mathcal{S}}(T,V)+S_{\text{b}}(T,V)=\left(  \frac{\partial
T\overline{\mathcal{S}}}{\partial T}\right)  _{\varphi_{\text{b}},V},
\]
where we have used (\ref{Basin_Complexity_Relation1}). Thus, the entropy of
the system is a sum of two contributions, which proves our statement (S4). The
derivative on the right-hand side in the above equation can also be written as
$-\left(  \partial F/\partial T\right)  _{\varphi_{\text{b}},V},$ which should
be identical to the derivative $-(\partial F/\partial T)_{V}$ from the
definition of $S.$ Thus,%
\[
(\partial F/\partial T)_{V}\equiv\left(  \partial F/\partial T\right)
_{\varphi_{\text{b}},V},
\]
which shows that the equilibrium free energy $F(T,V)$ is independent of the
basin free energy $\varphi_{\text{b}}$ $($at fixed $T).$ This should come as
no surprise since the equilibrium $\varphi_{\text{b}}$ is determined by the
condition
\begin{equation}
(\partial F/\partial\varphi_{\text{b}})_{T,V}\equiv(\partial\Phi
(\varphi,T,V)/\partial\varphi)_{T,V}=0. \label{EqF}%
\end{equation}

We can now include the kinetic energy part of the free energy to obtain the
complete free energy
\begin{align*}
\mathcal{F}(T,V)  &  \equiv-T\ln\mathcal{Z}(T,V)=\{E_{\text{b}}%
(T,V)+E_{\text{KE}}(T)\\
&  -T[S_{\text{b}}(T,V)+S_{\text{KE}}(T)]\}-T\overline{\mathcal{S}}(T,V).
\end{align*}
The quantity in the curly brackets represents what is conventionally called
the equilibrium vibrational free energy, even though one can be sure of its
vibrational nature only at low temperatures where we expect the projections
$d\Pi_{j}(E,V)$ of each contributing basin to be closed (for vibrations, the
basin must confine the motion in all directions, which may not be possible at
high temperatures). The last term is the contribution due to the complexity.

\subsubsection{Equilibrium Identifier Energy $\mathcal{E}(T,V)$}

The equilibrium basin identifier energy $\mathcal{E}(T,V)$\ is, by definition,
given by
\begin{equation}
\mathcal{E}(T,V)\equiv\sum_{j=1,N_{\text{B}}(V)}\mathcal{E}_{j}(V)\mathcal{P}%
_{j}(T,V). \label{AvPEM}%
\end{equation}
It is clear that this equilibrium value \emph{cannot} be expressed as a
derivative of the partition function. It is easily remedied. We add a
dimensionless parameter $\lambda$ that couples to the basin identifier energy
$\mathcal{E}_{j}$ so that there is an additional term $\lambda\mathcal{E}_{j}$
in the energy$,$ and define a modified PF%
\begin{equation}
Z(\lambda,T,V)\equiv\sum_{j=1,N_{\text{B}}(V)}e^{-\lambda\beta\mathcal{E}%
_{j}(V)}Z_{j}(T,V). \label{ModifiedPF}%
\end{equation}
The equilibrium $\mathcal{E}(T,V)$ is given by%
\[
\mathcal{E}(T,V)\equiv-T[(\partial\ln Z(\lambda,T,V)/\partial\lambda
)_{T,V}]_{\lambda=0}.
\]
In the limit $\lambda\rightarrow0,$ this modified PF reduces to the earlier PF
in (\ref{PF}). For $\lambda>0,$ basins with lower identifier energies are
weighted more than the higher ones, with the reverse true for $\lambda<0$. For
$\lambda=0,$ there is \emph{no} preference for any particular value of
$\mathcal{E}_{j};$ recall that $Z_{j}(T,V)$ does not depend on $\mathcal{E}%
_{j}.$\ It is obvious that the sum in (\ref{AvPEM}) can be limited to only
equilibrium basins for a macroscopic system. For equilibrium basins, all
allowed values of $\mathcal{E}_{j}$ are equally probable and will result in
the maximum spread $\overline{\Delta\mathcal{E}}$ in the possible values of
$\mathcal{E}_{j}$ under the summation in (\ref{AvPEM}). Thus, it appears that
the spread in the allowed values of $\mathcal{E}_{j}$ may very well be of the
order of $N,$ and not $\surd N.$\ In other words, the summand in (\ref{AvPEM})
is most probably not peaked sharply at some particular value of $\mathcal{E};$
see also Sect. IV(D) below. We hope to answer this question in future.

It is easy to show from (\ref{AvPEM}) that
\begin{equation}
T^{2}\frac{\partial\mathcal{E}(T,V)}{\partial T}=\sum_{j=1}^{N_{\text{B}}%
(V)}\Delta\mathcal{E}_{j}(V)\Delta E_{j}(T,V)\mathcal{P}_{j}(T,V),
\label{IdentifierCoefficient}%
\end{equation}
where $\Delta\mathcal{E}_{j}(V)\equiv\mathcal{E}_{j}(V)-\mathcal{E}(T,V),$
$\Delta E_{j}(T,V)\equiv E_{j}(T,V)-E(T,V),$ and $E(T,V)$ is the equilibrium
energy of the system. Thus, the derivative is a cross-correlation between two
fluctuations, $\Delta\mathcal{E}_{j}(V)$ and $\Delta E_{j}(T,V).$\ Since
cross-correlations do not usually have a fixed sign, there is no theoretical
reason for $\mathcal{E}(T,V)$ to be a monotonic increasing function of $T.$
This does not mean that the above derivative cannot be positive for many
physical systems.

We now show the expected result that the equilibrium free energy $F(T,V)$
cannot depend on the identifier energy explicitly by demonstrating that
\begin{equation}
(\partial F/\partial\mathcal{E})_{T,V}=0; \label{IdentifierFredom}%
\end{equation}
compare with (\ref{EqF}). For this, we use $\Phi(\varphi,T,V)\equiv
\varphi-T\mathcal{S}$ and obtain%
\[
(\partial\Phi/\partial\mathcal{E})_{T,V}=(\partial\varphi/\partial
\mathcal{E})_{T,V}-T(\partial\mathcal{S}/\partial\mathcal{E})_{T,V},
\]
which, from (\ref{DifferentialPhi}), reduces to%
\[
(\partial\Phi/\partial\mathcal{E})_{T,V}=(\partial\varphi/\partial
\mathcal{S})_{T,V}(\partial\mathcal{S}/\partial\mathcal{E})_{T,V}%
-T(\partial\mathcal{S}/\partial\mathcal{E})_{T,V}.
\]
In view of (\ref{Equilibriumphi}) at equilibrium, the right hand side is
identically zero, thereby proving (\ref{IdentifierFredom}).

\subsubsection{Monotonic $\mathcal{E}(T,V)$}

Because of the importance of different identifiers in different temperature
ranges, it is desirable to express thermodynamic quantities in terms of
$\mathcal{E}(T,V)$\ in these temperature ranges. Therefore, the question
arises whether it is possible to relate thermodynamic quantities
with\ $\mathcal{E}(T,V).$ However, as discussed above, these identifiers do
not determine the thermodynamics developed here; see (\ref{IdentifierFredom}),
because of which the relationship with $\mathcal{E}(T,V)$ will not be unique,
unless some particular scheme is used. We present a way to accomplish this
under the condition that $\mathcal{E}(T,V)$ is a monotonic function of $T$ in
its temperature range. We will, therefore, consider this case below. The
monotonic function $\mathcal{E}(T,V)$ of $T$ can be inverted to yield
\[
T\equiv T[\mathcal{E}(T,V),V].
\]
This will allow us to express any function of $T$ as a function of
$\mathcal{E}(T,V).$ For example, we can express $f_{j}(T,V)$ as a function
$f_{j}[\mathcal{E}(T,V),V]$ of $\mathcal{E}(T,V).$ We can similarly express
$\varphi_{j}(T,V)$ as a\ function $\varphi_{j}[\mathcal{E}(T,V),V]~$of
$\mathcal{E}(T,V).$ We can also express $\varphi_{\text{b}}(T,V),$ and
$\overline{\mathcal{S}}(T,V)$ as a function $\varphi_{\text{b}}[\mathcal{E}%
(T,V),V],$ and $\overline{\mathcal{S}}[\mathcal{E}(T,V),V],$ respectively, of
$\mathcal{E}(T,V).$

Let $N_{\text{BI}}(\mathcal{E},V)d\mathcal{E}$ represents the number of basins
in which the basin identifier has its energy in the range $\mathcal{E}$\ and
$\mathcal{E}+d\mathcal{E},$\ and introduce the corresponding microcanonical
entropy $S_{\text{BI}}(\mathcal{E},V)\equiv\ln N_{\text{BI}}(\mathcal{E}%
,V).$\ The $T$-dependent basin identifier entropy evaluated at $\mathcal{E}%
=\mathcal{E}(T,V)$
\begin{equation}
S_{\text{BI}}(T,V)\equiv S_{\text{BI}}[\mathcal{E}(T,V),V] \label{BIEntropy}%
\end{equation}
does not necessarily represent the entropy due to equilibrium basins for which
$\mathcal{E}=\mathcal{E}(T,V).$ The equilibrium basins satisfy the conditions
in (S2), but the basins enumerated by $S_{\text{BI}}(T,V)$\ have no such
conditions imposed on them. Moreover, the equilibrium basins have a wide range
of allowed basin identifier energies $\mathcal{E}_{j}$ because of $\lambda=0,$
as discussed above.\ Thus, the entropy due to equilibrium basins with
$\mathcal{E}=\mathcal{E}(T,V)$ is most probably very different from the
complexity $\overline{\mathcal{S}}[\mathcal{E}(T,V),V]$ obtained by the above
prescription from $\overline{\mathcal{S}}(T,V)$. It is clear that while the
basins counted by $N_{\text{BI}}(T,V)\equiv N_{\text{BI}}[\mathcal{E}(T,V),V]$
contain all the equilibrium basins with $\mathcal{E}=\mathcal{E}(T,V)$, they
also contain many other basins not satisfying (S2). Let $N_{\text{BI,NE}%
}(T,V)$ represent the latter non-equilibrium basins. Since not all equilibrium
basins in $\overline{\mathcal{N}}(T,V)$ have $\mathcal{E}=\mathcal{E}(T,V),$
the number of equilibrium basins $N_{\text{BI,E}}(T,V)$ in $N_{\text{BI}%
}(T,V)$\ is such that $N_{\text{BI,E}}(T,V)<$\ $\overline{\mathcal{N}}%
(T,V).$\ By definition, we have
\begin{equation}
N_{\text{BI}}(T,V)=N_{\text{BI,E}}(T,V)+N_{\text{BI,NE}}(T,V).
\label{BIEquality}%
\end{equation}
The current analysis cannot answer whether the second contribution is
thermodynamically insignificant compared to the first contribution in
(\ref{BIEquality}); see the discussion of (\ref{BasinEquality}) below. To have
$N_{\text{BI}}(T,V)=\overline{\mathcal{N}}(T,V)$\ at all temperatures would be
a precarious coincidence and hard to justify. Therefore, the sum
\[
S_{\text{BI}}(T,V)+S_{\text{b}}(T,V)
\]
will most probably not represent the equilibrium entropy $S(T,V)$ at all
temperatures$.$ For\ $\mathcal{E=}E_{\text{L}},$ $S_{\text{BI}}(T,V)$
represents the entropy $S_{\text{BM}}(T,V)$ introduced above in
(\ref{GoldsteinConfPF}). Thus, most probably%
\begin{equation}
S(T,V)\neq S_{\text{BM}}(T,V)+S_{\text{b}}(T,V) \label{BMEntropyPartition}%
\end{equation}
at all temperatures. In other words, $S_{\text{BI}}(T,V)$ is most probably not
equal to $\overline{\mathcal{S}}(T,V).$

As said above, the representation a thermodynamic quantity in terms of
$\mathcal{E}(T,V)$\ is not necessarily unique. Consider, for example$,$%
\ $\overline{\mathcal{S}}(T,V).$ There is an alternative way to express it in
terms of $\mathcal{E}(T,V)$ by recalling its fundamental definition
$\overline{\mathcal{S}}(T,V)\equiv\mathcal{S}[\varphi_{\text{b}}(T,V),T,V]$
given in $(\ref{InterbasinEntropy}).$ We express $\varphi_{\text{b}}(T,V)$ in
the form $\varphi_{\text{b}}(\mathcal{E},V)$ to obtain an alternative
expression $\mathcal{S}_{\varphi_{\text{b}}}(\mathcal{E},T,V)\equiv$
$\mathcal{S}[\varphi_{\text{b}}(\mathcal{E},V),T,V]$ for the complexity family
in which $\mathcal{E}(T,V)$ has been represented simply by $\mathcal{E}$. Let
us now consider the following derivative%
\begin{equation}
(\partial\overline{\mathcal{S}}/\partial\mathcal{E})_{T,V}\equiv
(\partial\mathcal{S}_{\varphi_{\text{b}}}(\mathcal{E},T,V)/\partial
\mathcal{E})_{T,V}=\beta(\partial\varphi_{\text{b}}/\partial\mathcal{E}%
)_{T,V}, \label{ComplexityIdentifier1}%
\end{equation}
which can be obtained from (\ref{DifferentialS}), and where we have also used
(\ref{Equilibriumphi}). We should recall that $\mathcal{E=E}(T,V)$\ above.

\subsubsection{Equiprobability Principle}

From (\ref{Prob}), and (\ref{FreeEnergy0}), we note that
\begin{equation}
\mathcal{P}_{j}(T,V)=\exp[-\mathcal{S}(T,V)], \label{Prob1}%
\end{equation}
for equilibrium basins, so that the system has \emph{equal} probability
$\exp[-\mathcal{S}(T,V)]$ to belong to any of the equilibrium basins in
$\mathcal{N}(\varphi,T,V)$. This proves (S3). The basins counted in
$\mathcal{N}(\varphi_{\text{b}},T,V)$ may have different $f_{j}(T,V)\equiv$
$\varphi_{\text{b}}(T,V)-$ $\mathcal{E}_{j},$ as there is no guarantee that
all these basins will have their basin identifier at the same energy. The
probability $\mathcal{P}_{j}(T,V)$ is determined only by $\varphi_{\text{b}%
}(T,V),$ the free energy of each of the equilibrium basins in
(\ref{FreeEnergy0}), which in turn determines $F(T,V)$ and $\overline
{\mathcal{S}}(T,V)$. Comparing with the standard Boltzmann equiprobability
principle, we can understand (\ref{Prob1}) by the following analogy (see also
\cite{Coluzzi}): the equilibrium basins and their free energies play the role
of microstates and their energies in the microcanonical ensemble; the basins
and their free energies play the role of microstates and their energies in the
canonical ensemble.

\section{Standard Picture}

We can rewrite the configurational PF in (\ref{ConfPF}) slightly differently
in a form, which is more standard and does not involve individual basins:%

\begin{equation}
Z(T,V)\equiv\int dEW(E,V)e^{-\beta E(V)}, \label{StandardPF}%
\end{equation}
where $W(E,V)dE$ represents the number of microstates or configurations
represented by the the slice $d\Sigma(E,V)$ with energy between $E$ and
$E+dE.$ We obtain the number of microstates in the slice $d\Sigma(V)$ by
considering the area of its projection $d\Pi(E,V)$ on $\Gamma(V)$:
$W(E,V)dE=\left\vert d\Pi(E,V)\right\vert .$ It is evident that
\begin{equation}
d\Pi(E,V)=\widetilde{\cup}_{j}d\Pi_{j}(E,V), \label{SliceUnion}%
\end{equation}
where the union is\ over only those basins that do not yield null projections;
these are the basins for which $E\notin\Delta_{j}E.$ The tilde over the union
denotes this restriction.\ Consequently, we have
\begin{equation}
W(E,V)=\widetilde{\sum}_{j=1,N_{\text{B}}(V)}\text{\ }W_{j}(E,V),
\label{ConfPartition}%
\end{equation}
where the tilde has the same meaning as in (\ref{SliceUnion}). Introducing the
microcanonical entropy $S(E,V)\equiv\ln$ $W(E,V),$ we conclude (see Remark 1)
that the PF in (\ref{StandardPF}) is dominated by the equilibrium energy
$E(T,V)$ for which the integrand is maximum, and gives the equilibrium state.
The equilibrium energy is the solution of
\begin{equation}
\left[  \partial S(E,V)/\partial E\right]  _{E=E(T,V)}=\beta,
\label{EquilibriumE}%
\end{equation}
which is obtained by the variation of the integrand at constant $T,V.$ At a
given temperature $\ T,$ only those microstates with energy in the range
$E(T,V)$, and $E(T,V)+dE$ determine the thermodynamics through the entropy
$S(T,V)\equiv S[E(T,V),V]$. The canonical entropy $S(T,V)$ along with $E(T,V)$
determine the equilibrium free energy $F(T,V)=E(T,V)-TS(T,V).$

As the kinetic energy is not a part of the energy $E,$ there is an upper bound
to the energy \cite{Note}. Therefore, the microcanonical entropy $S(E,V)$ near
this upper bound will be a decreasing function of $E.$ This region corresponds
negative temperatures according to (\ref{EquilibriumE}). Since $S(E,V)$ is an
increasing function near the lower end of the energy $E,$ there is expected to
be a maximum in $S(E,V)$ as a function of $E$. The rising branch of $S(E,V)$
corresponds to positive temperatures and it is this branch that is of interest
in this work.

\subsection{Connection with Landscape: Selectivity}

Using (\ref{IntrabasinFREE}) for the basin free energy $\varphi_{\text{b}%
}(T,V)$ in (\ref{FreeEnergy0}), and comparing with the above definition of the
free energy in the standard picture, we find that%
\begin{equation}
E(T,V)=E_{\text{b}}(T,V). \label{EnergyEquality}%
\end{equation}
The significance of this result is the following. All configurations
contributing to the canonical entropy $S(T,V)$ in the thermodynamic limit have
the same equilibrium energy equal to $E(T,V).$ The above equation
(\ref{EnergyEquality}) shows that these configurations, which can also be
classified as belonging to the different basins, as shown in
(\ref{ConfPartition}), not only have the same equilibrium basin energy
$E_{\text{b}}(T,V),$ but also have the same basin free energy $\varphi
_{\text{b}}(T,V);$ see (\ref{InterbasinEntropy}). From (\ref{IntrabasinFREE}),
this also implies that they all have the same basin entropy $S_{\text{b}%
}(T,V).$ Thus, the landscape thermodynamics is highly \emph{selective}. At a
given $T,$ only those basins contribute to the thermodynamics that have the
\emph{same} $E_{\text{b}}(T,V),S_{\text{b}}(T,V),$ and $\varphi_{\text{b}%
}(T,V).$ This proves our statement (S2) in totality.

\subsection{Equilibrium Basins}

Let $N_{\text{B}}(T,V)$\ denote the number of basins for which $E=E_{\text{b}%
}(T,V)\in\Delta_{j}E.$ We say that theses basins \emph{exist} at $T.$ Not all
these basins will satisfy all the conditions in (S2). The equilibrium basins
are the only basins that determine the thermodynamics and must satisfy (S2).
The number of these basins is $\overline{\mathcal{N}}(T,V)\equiv$
$\mathcal{N}(\varphi_{\text{b}},T,V).$ Thus, we expect $\overline{\mathcal{N}%
}(T,V)<N_{\text{B}}(T,V).$ The number of non-equilibrium basins $N_{\text{NE}%
}(T,V)$ that do not contribute to the thermodynamics of the system but exist
otherwise in the above sense is given by their difference:%
\begin{equation}
N_{\text{B}}(T,V)=\overline{\mathcal{N}}(T,V)+N_{\text{NE}}(T,V).
\label{BasinEquality}%
\end{equation}
Based on general arguments, there is no way to answer whether the second
contribution is thermodynamically insignificant compared to the first
contribution in (\ref{BasinEquality}). All we can be sure of is the following.
Because of (\ref{EntropyPartition}), we know for sure that the equilibrium
number of configurations $W(T,V)\equiv W[E(T,V),V]=\exp[S(T,V)]$ is
exhaustively coming from configurations in all the $\overline{\mathcal{N}%
}(T,V)$ equilibrium basins$:$%
\[
\overline{W}(T,V)=\overline{\mathcal{N}}(T,V)\exp[S_{\text{b}}(T,V)].
\]
The number of configurations coming from the $N_{\text{NE}}(T,V)$
non-equilibrium basins is thermodynamically insignificant. This result has no
bearing on the relative magnitude of the two terms in (\ref{BasinEquality}).
It is possible to have $N_{\text{NE}}(T,V)$ significantly larger than
$\overline{\mathcal{N}}(T,V)$ in (\ref{BasinEquality}) and still have the
contribution from the non-equilibrium basins to $W(T,V)$ insignificant
compared to $\overline{W}(T,V).$

\section{Discussion and Conclusions}

\subsection{Basin Connectivity and Percolation}

We have already noted, see (\ref{SliceUnion}), that the projection $d\Pi(E,V)$
is the union of different non-null components $d\Pi_{j}(E,V)$. A component may
have more than one disjoint pieces, such as the one from a basin in the form
of an inverted Mexican hat discussed earlier. Corresponding to some energy $E$
is the equilibrium temperature $T$ such that
\begin{equation}
E(T,V)=E. \label{TE-relation}%
\end{equation}
\ Because the heat capacity $C_{V}$\ is non-negative, $E(T,V)$\ is a monotonic
increasing function of $T$\ over the positive temperature range, which is what
we are interested in here. Thus, increasing(decreasing) $E$ is the same as
increasing(decreasing) $T.$\ 

The number of components at $E$ in the union in (\ref{SliceUnion}) is
$N_{\text{B}}(E,V)$\ introduced earlier. The shape and size of the components
and their number change with $E.$ The components $d\Pi_{j}(E,V)$\ cannot be
open; they are either connected together to form a cluster, or are isolated,
and must be closed (in the form of a ring). Two components are connected, when
there is no energy barrier separating them \cite{BarrierNote}. As we reduce
$E$ or $T,$ these components become disconnected due to the presence of an
energy barrier separating them. As Goldstein has pointed out \cite{Goldstein},
these energy barriers become very important at lower temperatures as we will
discuss further in the following. If a component is isolated, this will mean
that there is an energy barrier separating this basin from all other basins.
At very high energies ($E>E_{\text{CP}};$ see \cite{Note}) that are higher
than all energy barriers, all projections are open, but connected together. It
contains only a small number of basins containing particles near the walls.
These energies are not of any interest here. As the energy is reduced, some
energy barriers are cut by the surfaces $\widehat{\Sigma}(E)$ and
$\widehat{\Sigma}(E+dE),$ which produce isolated closed components and will
increase the number of components. In the same process, some of the components
disappear as their basin minima energy $E_{\text{L}}$ become higher than $E.$
This will reduce the number of components. Thus, we expect the number of
components to change with $E.$

For a macroscopic system, the above idea of basin connectivity in the
$3N$-dimensional phase space must be phrased in a way that makes statistical
sense. A basin $i$ is connected to a basin $j$, if the measure of their common
opening (along which there is absence of any barrier) is a non-zero fraction
of the projection measure $\left\vert d\Pi_{i}(E,V)\right\vert .$ We will call
such openings \emph{relevant}. \ The non-zero fraction of the projection
measure will add a non-zero entropic contribution to the barrier height. (It
is almost impossible for the system to find a barrier along one of the
directions out of macroscopically large $3N$ directions over any finite, no
matter how long, time; the entropic contribution for such a barrier is zero.)
Thus, it is the free energy , and not the energy that becomes relevant for
barrier hopping. A relevant opening will have a non-zero probability
$\mathcal{P}_{ij}$ of a jump $B_{i}\rightarrow B_{j}$ from the $i$-th basin to
the $j$-th basin. Openings that are zero fraction of the projection will be
called \emph{irrelevant }as $\mathcal{P}_{ij}=0$. \ It is in the sense of a
relevant opening that the two basins will be considered connected below.
Connected basins allow the system to probe them without encountering any
energy barriers. In the following, we will make distinctions between two kinds
of barriers \cite{BarrierNote}. Barriers whose heights above $E(T)$ are of the
order of $T$ can be overcome due to thermal fluctuations ($\overline{\Delta
E}=T\surd C_{V}$) and the time required for the jump $B_{i}\rightarrow B_{j}$
is less than or equal to the microscopic relaxation time. If the interest is
to study dynamics at a much longer range of time, we can treat these barriers
as not different from an opening. The barriers whose heights are much higher
than $T$ will give rise to \emph{activation processes} requiring much longer
times to jump over the barriers. These \emph{high barriers} are the ones we
consider in the following. The \emph{most probable barriers} are those
barriers that have the maximum probability to be overcome at a given $T.$ The
barriers whose heights above $E(T)$ are of the order of $T$ will be treated as
openings; thus, they are suppressed \cite{BarrierNote}.

It is obvious that the physics of a continuum model requires the system in the
\emph{diffusive regime} to probe various configurations that continuously
transform into each other, as long as we are far away from any phase
transition or we are in a metastable state obtained by avoiding a transition
like the crystallization. Therefore, we expect various basins to be
\emph{continuously connected} to each other in the sense that there is at
least one continuous passage from one basin to another through intermediate
basins, without any energy barriers to overcome, so that the phase point can
move continuously from one basin to the other. This will certainly be the case
at high enough energies (that correspond to high enough temperatures). If it
were not the case, the system will have to overcome energy barriers using
activation processes because of which, it will "jump" from the equilibrium
configuration in one basin into another equilibrium configuration of the other
basin (with energy barriers between them); the configuration of the new basin
is different from that in the old basin in that the new configuration is not
in the immediate vicinity of the old one. The high-temperature situation will
correspond to all the components in the projection $d\Pi(E,V)$ being connected
together as one connected cluster. We can visualize this situation as the
\emph{percolation} of the basin slice projections $d\Pi_{j}(E,V)$ in one
macroscopic cluster, called the \emph{percolating cluster }$\mathcal{D}(E,V)$
that spans the entire space $\Gamma(V)$. Here, the percolation occurs between
components $d\Pi_{j}(E,V),$ all of different shapes and sizes.

The entropy of the system $S(E,V)$ is determined by the percolating cluster
$\mathcal{D}(E,V)$\ at high enough energies. There are other finite clusters
or isolated components in addition to $\mathcal{D}(E,V).$ But not all
components and clusters are thermodynamically significant and contribute to
the entropy. Only those components contribute that have the same
$E(T,V),S_{\text{b}}(T,V),$ and $\varphi_{\text{b}}(T,V).$ Because of the
continuity in the diffusive regime, we expect these components to be part of
$\mathcal{D}(E,V).$\ The number of such components is given by $\mathcal{N}%
(\varphi_{\text{b}},T,V).$ At high energies, $\mathcal{D}(E,V)$ contains
\emph{disordered} configurations pertinent for the liquid state.\ The
percolating cluster allows the system to probe all of the connected
equilibrium basins that occur in the diffusive regime.

As we reduce the energy, some of the basin components (i) disappear since they
no longer exist ($E<E_{\text{L}}$), (ii) appear since their energy range
contains the energy $E$ ($E_{\text{L}}<E<E^{\text{H}}$)$,$ or (iii) begin to
get disconnected from others due to the emergence of high energy barriers, but
the percolating cluster $\mathcal{D}(E,V)$ persists in $d\Pi(E,V)$ although it
is of a different size. It may happen that some basin components get
completely \emph{isolated} from each other and the percolating cluster. It is
highly plausible that in the interesting range of energies, $\mathcal{D}(E,V)$
will decrease in size as $E$ is reduced. However, the entropy of the system is
still determined by the percolating cluster, at least at higher energies that
correspond to temperatures much higher than the melting temperature
$T_{\text{M}}$. Close to but still above $T_{\text{M}},$ another cluster
$\mathcal{C}(E,V)$ containing \emph{ordered} configurations relevant for the
crystal state begins to form, and begins to compete with the disordered
cluster $\mathcal{D}(E,V)$ as far as their entropies are concerned. At
$T_{\text{M}},$ there is a transition to the crystal, and the thermodynamics
is now controlled by $\mathcal{C}(E,V).$

As we are interested in supercooled states, we will preclude the basins that
determine $\mathcal{C}(E,V).$ This gives rise to the restricted PF formalism
to study metastability \cite{Penrose}. Consequently, we continue to consider
$\mathcal{D}(E,V)$ as we reduce the energy. At lower energies or temperatures,
the cluster $\mathcal{D}(E,V)$ will be fragmented enough so that the
percolation ceases to occur mainly because the number of components has
drastically decreased, and all we have are isolated components or clusters of
components. The disconnected components will then force the system to undergo
activation-assisted jumps between different basins and will force the system
to undergo the dynamic glass transition. Thus, the picture strongly suggests a
tantalizing possibility that the above percolation transition has direct
relevance for the mode-coupling transition. We are working on this issue and
hope to report soon in a separate publication.

\subsection{Free energy distribution}

The basins that contribute to the thermodynamics are enumerated by
$\mathcal{N}(\varphi,T,V).$ All these basins have the same equilibrium energy,
entropy and the free energy, as discussed above. Other basins containing
configurations of energy $E(T,V),$ but not satisfying the other conditions do
not contribute, even though they may be infinitely many, as discussed above.
Some of them may even have their free energies $\widetilde{\varphi}(T,V)$
$<$%
$\varphi_{\text{b}}(T,V);$ they still will not determine the thermodynamics
because the corresponding free energy $\widetilde{F}(T,V)$ due to basins with
free energy $\widetilde{\varphi}(T,V)$ is higher than the equilibrium free
energy $F(T,V)$ discussed above in (\ref{FreeEnergy0}). Such a possibility
($\varphi_{j}=\widetilde{\varphi}$
$<$%
$\varphi_{\text{b}}$) has an interesting consequence. The probability for
probing one of the nonequilibrium basins is%
\begin{equation}
\mathcal{P}_{j}(T,V)=\exp[-\mathcal{S}-\beta(\widetilde{\varphi}%
-\varphi_{\text{b}})], \label{Prob2}%
\end{equation}
and is higher than the probability to be in an equilibrium basin; here,
$\mathcal{S}$ is the equilibrium complexity $\mathcal{S}(\varphi_{\text{b}%
},T,V).$ There is nothing wrong with it. Let $\exp[\widetilde{\mathcal{S}%
}(\widetilde{\varphi},T,V)]$ denote the number of \ nonequilibrium basins with
the basin free energy $\widetilde{\varphi}.$ Then, the probability
$\widetilde{\mathcal{P}}(T,V)$ to be in any of these basins is obtained by
multiplying this number by $\mathcal{P}_{j}(T,V)$ above, which yields%
\[
\widetilde{\mathcal{P}}(T,V)=\exp[-\beta(\widetilde{F}-F)],
\]
which surely vanishes for a macroscopic system since $F$ is the lowest free
energy. Thus, we can neglect the nonequilibrium basins in thermodynamic
considerations, even though they are important for studying transition rates
from one state to another, which we discuss below.

\subsection{Transitions among Basins}

We have already argued that all equilibrium basins are equally probable; see
(\ref{Prob1}). This is nothing but the Boltzmann hypothesis of microstates
\cite{Landau}; the only difference is that the microstates are now replaced by
the basins, and the microstate energies $E$ are replaced by the basin free
energies $\varphi.$ Since there is no physical interactions between basins,
each basin can be thought of as representing a "microstate" of the system.
Thus, the emergence of the principle of equiprobability of basins is not
surprising. What is surprising is that this principle is valid even if we
consider a restricted PF in which we disallow certain configurations (or
basins) like those relevant for crystallization. Such a PF is used to study
metastable states like supercooled liquids
\cite{Goldstein,Goldstein1,Schulz,Penrose}. Thus, the above principle gives
rise to a "restricted" \emph{ergodic behavior} within the restricted PF
formalism. With this analogy, we can now begin to understand the
time-evolution of the system, since it is well-known \cite{Landau} that the
principle of equiprobability is equivalent to the principle of detailed
balance or Liouville theorem: If states (or basins) $B_{1}$ and $B_{2}$ have
the same (free) energy, then the probability $\mathcal{P}_{12}$ per unit time
of a transition from $B_{1}\rightarrow B_{2}$ and the probability
$\mathcal{P}_{21}$ per unit time of the reverse transition from $B_{2}%
\rightarrow B_{1}$ are the same:%
\[
\mathcal{P}_{12}=\mathcal{P}_{21}.
\]
The rate at which the transition $B_{1}\rightarrow B_{2}$ takes place is%
\[
\mathcal{R}_{12}=\mathcal{P}_{1}\mathcal{P}_{12},
\]
where $\mathcal{P}_{1}$ is the probability that the system explores the basin
$B_{1};$ we have suppressed the arguments $T,V$ for simplicity. It is clear
that $\mathcal{R}_{12}=\mathcal{R}_{21}.$

It should be noted that $\mathcal{P}_{ij}$ for transitions $B_{i}\rightarrow
B_{j}~$need not all be the same for all choices of $i$ and $\ j.$ Indeed, some
of them may be zero. Assume that at a given instance, the system is probing a
basin $B_{1}$ and has a very small probability of getting into some basin
$B_{0},$ because there is only one pathway through intermediate (equilibrium
and nonequilibrium) basins $B_{2},B_{3},B_{4},....$ to $B_{0}.$ While there
may be an appreciable probability for the transition $B_{1}\rightarrow B_{2},$
$B_{1}\rightarrow B_{3},$ etc. because of a large number of pathways
connecting $B_{1}\ $to $B_{2},$ $B_{1}\ $to $B_{3},$ etc., the probability of
transition $B_{1}\ $to $B_{0}$ remains very small. This also implies that the
probability of the reverse transition $B_{0}\ $to $B_{1}$ is also small as
there is only one path connecting $B_{0}\ $to $B_{1}.$ It may also happen that
the probability of transition to other basins from $B_{0}$ is very small. This
will mean not only that it is highly unlikely for the system to get into the
basin $B_{0},$ but also for it to escape $B_{0}.$ In no way does this violate
the principle of equiprobability. What this situation corresponds to is this:
once the system begins to probe the basin $B_{0},$ it stays there for a long
time before escaping. The amount of time the system probes the basin $B_{0}$
is still the same as the amount of time it spends in any other basins
$B_{1},B_{2},B_{3},$ etc., which is what the principle of equiprobability
implies \cite{Landau}.

The rate $\overset{\centerdot}{\mathcal{P}}_{i}$ at which $\mathcal{P}_{i}$
changes with time is given by
\[
\overset{\centerdot}{\mathcal{P}}_{i}\equiv\sum_{j}\mathcal{P}_{j}%
\mathcal{P}_{ji}-\mathcal{P}_{i}\sum_{j}\mathcal{P}_{ij},
\]
where we must remember that $\mathcal{P}_{ji}=\mathcal{P}_{ij}.$\ In general,
the transition probability $\mathcal{P}_{ij}$ satisfies%
\begin{equation}
\mathcal{P}_{ij}\equiv1-\sum_{k\neq j}\mathcal{P}_{ik}. \label{BasinProb}%
\end{equation}
The above situation in which $B_{0}$\ has only a few escape routes to other
basins becomes more probable as the temperature is reduced. The value of
$\mathcal{P}_{0j}$ depends, not on the free energy heights of the energy
barriers\ but on their ratios with $T.$ Thus, at low temperatures,
$\mathcal{P}_{0j}$ can be very small for two disjoint basin components. That
does not mean that the system will be trapped in $B_{0},$ unless there are
only a few basins (thermodynamically insignificant in number) in the sum in
(\ref{BasinProb}) so that $\mathcal{P}_{0}\cong1.$ If there is a large number
of basins (thermodynamically significant in number) in the sum in
(\ref{BasinProb}), $\mathcal{P}_{0}<1$\ and the system will be able to escape
to some other basins. It is quite clear that the probability of staying in a
given basin increases as $T$ is reduced, mainly because there are only a few
basins left in the system at that temperature. This is the mechanism behind
the glass transition.

To study the trapping of the system to a single basin $B_{0}$ at low
temperatures, we can follow the standard idea of \emph{symmetry breaking} in
condensed matter and apply a symmetry breaking bias $\lambda_{0}(>0)$ to make
the free energy of $B_{0}$ lower than the free energies of other equiprobable
basins: $F_{0}(T,V)\equiv F(T,V)-\lambda_{0}.$ This will single out the basin
$B_{0}$ from all equiprobable basins. At the end of the calculation, we can
take the limit $\lambda_{0}\rightarrow0.$

\subsection{ Equilibrium identifier Energy $\mathcal{E}(T,V)$}

Let us introduce the \emph{mean} identifier energy
\begin{equation}
\widehat{\mathcal{E}}(\varphi,T,V)=\frac{1}{\mathcal{N}(\varphi,T,V)}%
\sum_{j=1,\mathcal{N}(\varphi,T,V)}\mathcal{E}_{j}(V)
\label{MeanIdentifierEnergy}%
\end{equation}
over all basins with the same free energy $\varphi.$ These need not be the
equilibrium basins. Furthermore, they not all have the same $\mathcal{E}%
_{j}(V).$ Using (\ref{Prob2}) in (\ref{AvPEM}), we can write%

\[
\mathcal{E}(T,V)=\sum_{j=1,N_{\text{B}}(V)}\mathcal{E}_{j}(V)\exp
[-\mathcal{S}-\beta(\varphi_{j}-\varphi_{\text{b}})].
\]
Let $\exp[\mathcal{S}(\widetilde{\varphi},T,V)]$ denote the number of basins
with the basin free energy $\widetilde{\varphi}.$ Grouping basins according to
their free energy $\widetilde{\varphi},$ and introducing $\widetilde
{F}(\widetilde{\varphi},T,V)=\widetilde{\varphi}-T\mathcal{S}(\widetilde
{\varphi},T,V),$ we can write%
\[
\mathcal{E}(T,V)=\sum_{\widetilde{\varphi}}\widehat{\mathcal{E}}%
(\widetilde{\varphi},T,V)\exp[-\beta(\widetilde{F}-F)].
\]
Since $F$ is the lowest free energy corresponding to $\widetilde{\varphi
}=\varphi_{\text{b}},$ we can replace the sum over $\widetilde{\varphi}$ by
the most dominant term corresponding to $\widetilde{\varphi}=\varphi
_{\text{b}},$ i.e. $\widetilde{F}=F.$ This finally gives
\[
\mathcal{E}(T,V)=\widehat{\mathcal{E}}[\varphi_{\text{b}}(T,V),T,V].
\]
The sum in (\ref{MeanIdentifierEnergy}) for $\widetilde{\varphi}%
=\varphi_{\text{b}}$ is over the equilibrium basins counted in $\mathcal{N}%
(\varphi_{\text{b}},T,V)$. The equilibrium identifier energy must be
extensive. The range $\mathcal{R(}T\mathcal{)}$ over which $\mathcal{E}_{j}$
varies in (\ref{MeanIdentifierEnergy}) will depend on the system, and it is
conceivable that $\mathcal{R(}T\mathcal{)}/N$ is not \emph{zero} in the
thermodynamic limit, as we discussed above. In this case, the free energy
$f_{j}(T,V)$\ will not be the same for all equilibrium basins. However,
Goldstein \cite{Goldstein} intuitively argues that $\mathcal{R(}%
T\mathcal{)}/N$ should vanish for basin minima as the identifier, and that the
basin minima have a sharp distribution about its equilibrium. If this is
correct, it would imply that all equilibrium basins also have the same average
minima in additions to the conditions in (S2). This will imply that all basins
are identical in the thermodynamic sense (though not necessarily in the
topological sense). All we can say is that our analysis is completely
oblivious of the basin minima or other identifier energies $\mathcal{E}_{j}$.

Going back to the modified PF in (\ref{ModifiedPF}), we observe that as
$\lambda\rightarrow0,$ there is no bias whether lower identifier energies are
weighted more ($\lambda>0$) or higher identifier energies ($\lambda<0$) are
weighted. However, there will in general be some topological restrictions. For
example, basins with $E_{\text{L}}\mathcal{>}E(T,V)$ or $E^{\text{H}%
}\mathcal{<}E(T,V)$ are not allowed, but this puts no strong restriction on
the range $\mathcal{R(}T\mathcal{)}$. For example, all basins with
$E_{\text{L}}\mathcal{<}E(T,V)$ regardless of how low they are will contribute
to $E_{\text{L}}(T,V),$ as long as the corresponding basin free energy is
equal to $\varphi_{\text{b}}(T,V).$ Thus, it is not clear whether
$\mathcal{R(}T\mathcal{)}/N$ is really zero in the thermodynamic limit.

\subsection{Complexity as Free Energy?}

The complexity appears in an additive fashion in the free energy, which
suggests the identification of $\mathcal{N}(\varphi,T,V)$ as some kind of a
configurational state PF $Z_{\text{CS}}(T,V)$ [think of $Z_{\text{CS}}(T,V)$
as the factor $Z_{\text{BM}}(T,V)$ in (\ref{GoldsteinPart})]$,$ even though
the former truly determines an entropic quantity $\mathcal{S}(\varphi,T,V)$,
called complexity; see (\ref{EntropyPartition}). Usually, the microcanonical
entropy in statistical mechanics is a natural function of energy, and not
temperature \cite{Landau}, whereas $\mathcal{N}(\varphi,T,V)$ is a function
the temperature. Despite this, we now show that it is not proper to think of
the above identification $Z_{\text{CS}}(T,V)\equiv\mathcal{N}(\varphi,T,V)$
for the following reason. The identification will yield the free energy
$F_{\text{CS}}(T,V)\equiv-T\ln Z_{\text{CS}}(T,V)=-T\overline{\mathcal{S}},$
from which we can compute the corresponding entropy [$-(\partial F_{\text{CS}%
}(T,V)/\partial T)_{V}$], and energy [$E=F+TS$]%
\begin{align*}
S_{\text{CS}}(T,V)  &  =\overline{\mathcal{S}}+T\left(  \partial
\overline{\mathcal{S}/}\partial T\right)  _{V},\\
E_{\text{CS}}(T,V)  &  =T^{2}\left(  \partial\overline{\mathcal{S}/}\partial
T\right)  _{V}.
\end{align*}
We now use this in (\ref{FreeEnergy0}), and obtain
\begin{align*}
F(T,V)  &  =E_{\text{b}}(T,V)+E_{\text{CS}}(T,V)\\
&  -T[S_{\text{b}}(T,V)+\overline{\mathcal{S}}(T,V)+T\left(  \partial
\overline{\mathcal{S}/}\partial T\right)  _{V}].
\end{align*}
This form of the free energy will give for the energy the first two terms,
which is not correct; see (\ref{EnergyEquality}). Similarly, the quantity in
the square rackets also is not the correct entropy of the system; see
(\ref{EntropyPartition}). Hence, complexity should not be considered as a free
energy. Its proper identification is with the entropy. This justifies using
the complexity as the entropy associated with configurational states. This
entropy is neither the entropy $S_{\text{CS}}(T,V)$ nor the entropy
$S_{\text{BI}}(T,V)$ due to any basin identifier. The entropy associated with
the configurational states advocated by Goldstein truly represents the entropy
associated with the basins of a given free energy $\varphi,$ and not a given
BM energy. Moreover, the configurational states of Goldstein should be
identified as the basins of a given free energy $\varphi.$

\subsection{Extension to NTP-Ensemble}

For the $NTP$-ensemble, we need to integrate $\mathcal{Z(}T,V)\exp(-\beta
PV)dV,$ see the PF in (\ref{TotalPF}), over the volume, so that the role of
$V$ in the above discussion is played by the equilibrium volume $\overline
{V}(T,P)$ and should be replaced by it$.$ All the above formulas remain valid
if we replace $V$ by either $P$ or $\overline{V}(T,P)$ in all thermodynamic
quantities$.$ (It is for this reason that we have explicitly shown $V$ in all
the formulas above.) For example, the basin energy is represented as a
function $E_{j}(T,P)$ or $E_{j}[T,\overline{V}(T,P)].$ The same is true of the
basin free energy $\varphi,$ which now represents the Gibbs free energy. The
complexity is to be treated as a function of $\varphi,T,$ and $P$ or
$\overline{V}(T,P).$

The PE landscape is now defined in a $(3N+2)$-dimensional landscape
space$~\Lambda_{N}$ in which $V$ is added as an additional axis. The PE
surface is now a $(3N+1)$-dimensional surface $\Sigma$. Since the
configuration space $\Gamma(V)$ requires a particular volume $V$ for its
definition$,$ we need to project $\Sigma$ for a particular value of $V.$ Thus,
we need to deal with the same sort of projection $\Pi(V)$ in the configuration
space as before.\ The slice $d\Sigma$ also needs to be projected for a
particular value of $V.$ Thus, we determines the number of configurations
$W(E,V)$ for a given $\ E,$ and $V.$ In equilibrium, corresponding to\ $E,$
and $V$ exist $T,$ and $P$ such that $E(T,P)=E,$ and $V(T,P)=V.~$All the
conclusions above remain valid.

\subsection{Summary}

In summary, we have investigated the landscape description at all
temperatures, even though the picture was originally introduced by Goldstein
as valid and useful only at low tempertures. We have succeded in validating
the picture at any temperature. Thus, the results we present here are not
restricted to any particular temperature. We have shown that the basins
represent the configurational states that were introduced by Goldstein. These
basins and their free energy $\varphi$ play the role of microstates and their
energy $E$ in the microcanonical ensemble, and are governed by a similar
equiprobability principle, which extends the Boltzmann equiprobability
principle for microstates. The corresponding entropy due to the number of
basins of a given free energyy is called the complexity \cite{Parisi}, and
correctly represents an entropy and not a free energy. We have shown that the
entropy is a sum of complexity and intrabasin entropy. We also conclude that
all equilibrium basins at a given temperature not only have the same
equilibrium free energy $\varphi_{\text{b}}(T,V),$ but also have the same
energy $E(T,V),$ and the same intrabasin entropy $S_{\text{b}}(T,V).$ Because
of the relevance of equilibrium basin identifiers in different temperature
ranges,we have provided a procedure to express thermodynamic quantities as a
function of equilibrium basin identifier energy $\mathcal{E}(T,V)$ since the
equilibrium free energy does not explicitly depend on $\mathcal{E}(T,V).$ Our
analysis provides justifications for two of the conjectures made by Goldstein
but with slight modifications. The equiprobability theorem allows to draw some
conjectural conclusions about the dynamics and the role of landscape
projection components in the phase space. We discuss the percolation of basin
projection components and speculate about the possible connection of the
disappearance of percolation with the dynamic glass transition. This topic is
being investigated at present and we hope to report on it in near future.

\bigskip\


\begin{thebibliography}{99}
\bibitem {Note}The potential energy when two or more particles overlap is
infinitely large. To avoid this infinity, we will assume that the centers of
mass of two particles are not allowed to be less than some distance $d.$ Let
$E_{d}>0$ be the potential energy between two particles at this distance. The
maximum energy when all the particles are at a distance $d$ from each other is
$\allowbreak N\left(  N-1\right)  E_{d}/2,$ but such an event is very rare.
Even the probability of having a large but finite numbers of particles being
at a distance$\ d$ from each other is very small. Hence, we will assume that
the highest allowed energy due to close proximity is $E_{\text{CP}}=cNE_{d},$
where $c$ is some small and non-zero constant. In this case, some of the
basins may have $E_{j}^{\text{H}}=$ $E_{\text{CP}}$. All other basins are
expected to have their $E_{j}^{\text{H}}<$ $E_{\text{CP}}.$
\bibitem {BarrierNote}The thermal fluctuations have been neglected in our
analysis so far. These fluctuations will allow the system to spontaneously hop
over barriers whose heights are of the order of $T.$ Recall that associated
with each energy $E$ in the landscape, there exists a unique temperature $T$
through (\ref{TE-relation})$.$ These spontaneous fluctuation jumps determine
some microscopic time scale of the system, and should be distinguished from
activation processes in which the system will jump over a much higher barrier
requiring much larger time scales, reaching in some cases to an astronomical
scale. To study these activation processes, it is advisable to suppress
barriers in the landscape whose heights are less than or of the order of $T,$
leaving behind a modified landscape that is probably more useful.
\end{thebibliography}
\end{document}